\documentclass[12pt]{article}
\usepackage{amssymb}

\usepackage{amsmath}
\usepackage[unicode,bookmarks,bookmarksopen,bookmarksopenlevel=2,colorlinks,linkcolor=blue,citecolor=green]{hyperref}

\def\comment#1{}
\input{tcilatex}

\begin{document}

\title{Transformations of integrable hydrodynamic chains and their
hydrodynamic reductions}
\author{Maxim V. Pavlov}
\date{}
\maketitle

\begin{abstract}
Hydrodynamic reductions of the hydrodynamic chain associated with
dispersionless limit of 2+1 Harry Dym equation are found by the Miura type
and reciprocal transformations applied to the Benney hydrodynamic chain.
\end{abstract}

\tableofcontents

\vspace{1cm}

\textit{keywords}: hydrodynamic chain, reciprocal transformation, Miura type
transformation, hydrodynamic reduction.


\section{Introduction}

Solutions of integrable hydrodynamic chains and 2+1 quasilinear systems can
be found by the method of hydrodynamic reductions (see, for instance, 
\textbf{\cite{Fer+Kar}}, \textbf{\cite{Gib+Tsar}}). Suppose some
hydrodynamic chain is given. Then one can seek hydrodynamic reductions for
this hydrodynamic chain by the aforementioned method of hydrodynamic
reductions or by $\bar{\partial}$ approach (see \textbf{\cite{Bogdan}}), for
instance. Thus, hydrodynamic reductions of a dispersionless limit of 2+1
Harry Dym equation can be found directly by the Hamiltonian approach applied
for the Kupershmidt hydrodynamic chain (see \textbf{\cite{Maks+Kuper}} and 
\textbf{\cite{Maks+Hamch}}) or by $\bar{\partial}$ approach (see \textbf{%
\cite{Wu}}). However, in this paper we concentrate on re-calculation of
hydrodynamic reductions of hydrodynamic chains related by Miura type and
reciprocal transformations. It means that we use already known hydrodynamic
reductions of the Benney hydrodynamic chain (see \textbf{\cite{Bogdan}}, 
\textbf{\cite{Krich}}, \textbf{\cite{Zakh}}) for reconstruction of
hydrodynamic reductions for a dispersionless limit of 2+1 Harry Dym equation.

Thus, this paper is devoted to the construction of hydrodynamic reductions
common for 2+1 quasilinear systems (dKP, dmKP, 2+1 Harry Dym) related by the
Miura type and reciprocal transformations (see \textbf{\cite{Chang}}, 
\textbf{\cite{Chang+Tu}}, \textbf{\cite{Chen+Tu}}, \textbf{\cite{Rogers}}, 
\textbf{\cite{Shaw+Tu}}).

Let us recall (see \textbf{\cite{Gibbons}}) that the Benney hydrodynamic
chain (see \textbf{\cite{Benney}})%
\begin{equation}
A_{t}^{k}=A_{x}^{k+1}+kA^{k-1}A_{x}^{0}\text{, \ \ \ \ \ \ \ }k=0,1,2,...
\label{1}
\end{equation}%
satisfies the Gibbons equation%
\begin{equation}
\lambda _{t}-\mu \lambda _{x}=\frac{\partial \lambda }{\partial \mu }\left[
\mu _{t}-\partial _{x}\left( \frac{\mu ^{2}}{2}+A^{0}\right) \right] ,
\label{2}
\end{equation}%
where the equation of the Riemann mapping is given by the asymptotic series%
\begin{equation}
\lambda =\mu +\frac{A^{0}}{\mu }+\frac{A^{1}}{\mu ^{2}}+\frac{A^{2}}{\mu ^{3}%
}+...  \label{3}
\end{equation}%
The inverse asymptotic series%
\begin{equation*}
\mu =\lambda -\frac{\mathbf{H}_{0}}{\lambda }-\frac{\mathbf{H}_{1}}{\lambda
^{2}}-\frac{\mathbf{H}_{2}}{\lambda ^{3}}-...
\end{equation*}%
yields an infinite series of polynomial conservation law densities $\mathbf{H%
}_{k}(A^{0},A^{1},...,A^{k})$, where the generating function of conservation
laws is given by%
\begin{equation}
\mu _{t}=\partial _{x}\left( \frac{\mu ^{2}}{2}+A^{0}\right) .  \label{zak}
\end{equation}%
Since the transformation $\mathbf{H}_{k}=\mathbf{H}%
_{k}(A^{0},A^{1},...,A^{k})$ is invertible, the Benney hydrodynamic chain
can be written in the conservative form%
\begin{equation}
\partial _{t}\mathbf{H}_{0}=\partial _{x}\mathbf{H}_{1}\text{, \ \ \ \ }%
\partial _{t}\mathbf{H}_{n}=\partial _{x}\left( \mathbf{H}_{n+1}-\frac{1}{2}%
\overset{n-1}{\underset{k=0}{\sum }}\mathbf{H}_{k}\mathbf{H}_{n-1-k}\right) 
\text{, \ \ \ }n=1,2,...  \label{cons}
\end{equation}

Lets us apply the generating function of the Miura type transformations (see 
\textbf{\cite{Kuper}})%
\begin{equation*}
\mu =p+B^{0}
\end{equation*}%
to the above series (see also \textbf{\cite{Chang+Tu}}, \textbf{\cite%
{Shaw+Tu}}). A deformation of the Riemann mapping determined by the
asymptotic series%
\begin{equation}
\lambda =p+B^{0}+\frac{B^{1}}{p}+\frac{B^{2}}{p^{2}}+\frac{B^{3}}{p^{3}}+...
\label{5}
\end{equation}%
satisfies the modified Gibbons equation%
\begin{equation}
\lambda _{t}-(p+B^{0})\lambda _{x}=\frac{\partial \lambda }{\partial p}\left[
p_{t}-\partial _{x}\left( \frac{p^{2}}{2}+B^{0}p\right) \right] ,  \label{6}
\end{equation}%
where a dynamics of the coefficients $B^{k}$ is given by the modified Benney
hydrodynamic chain%
\begin{equation}
B_{t}^{k}=B_{x}^{k+1}+B^{0}B_{x}^{k}+kB^{k}B_{x}^{0}\text{, \ \ \ \ \ \ }%
k=0,1,2,...  \label{7}
\end{equation}

\textbf{Remark}: The comparison of two asymptotic series (\textbf{\ref{3}})
and (\textbf{\ref{5}})%
\begin{equation*}
p+B^{0}+\frac{B^{1}}{p}+\frac{B^{2}}{p^{2}}+\frac{B^{3}}{p^{3}}+...=p+B^{0}+%
\frac{A^{0}}{p+B^{0}}+\frac{A^{1}}{(p+B^{0})^{2}}+\frac{A^{2}}{(p+B^{0})^{3}}%
+...
\end{equation*}%
yields explicit polynomial Miura type transformations $%
A^{k}(B^{0},B^{1},...,B^{k+1})$, $k=0,1,2,...$

The relationship between 2+1 Harry Dym equation and the KP equation was
established in \textbf{\cite{Rogers}} (see also \textbf{\cite{Chang}}). Let
us recall this link between corresponding hydrodynamic chains (see \textbf{%
\cite{Chen+Tu}}). The hydrodynamic chain connected with a dispersionless
limit of 2+1 Harry Dym equation is (see, for instance, \textbf{\cite{Chang}})%
\begin{equation}
C_{y}^{k}=(C^{-1})^{2}C_{z}^{k+1}+(k+1)C^{k+1}C^{-1}C_{z}^{-1}\text{, \ \ \
\ \ \ }k=-1,0,1,2,...  \label{g}
\end{equation}%
A deformation of the Riemann mapping%
\begin{equation}
\lambda =C^{-1}q+C^{0}+\frac{C^{1}}{q}+\frac{C^{2}}{q^{2}}+\frac{C^{3}}{q^{3}%
}+...  \label{8}
\end{equation}%
is given by the Gibbons equation%
\begin{equation}
\lambda _{y}-q(C^{-1})^{2}\lambda _{z}=\frac{\partial \lambda }{\partial q}%
\left[ q_{y}-\partial _{z}\left( \frac{q^{2}(C^{-1})^{2}}{2}\right) \right] .
\label{9}
\end{equation}%
This hydrodynamic chain has the first conservation law%
\begin{equation*}
\partial _{y}\frac{1}{C^{-1}}=\partial _{z}(-C^{0}).
\end{equation*}%
Under the reciprocal transformation%
\begin{equation}
dx=\frac{1}{C^{-1}}dz-C^{0}dy\text{, \ \ \ \ \ }dt=dy  \label{soh}
\end{equation}%
the hydrodynamic chain (\textbf{\ref{g}}) reduces to the hydrodynamic chain (%
\textbf{\ref{7}}); the Gibbons equation (\textbf{\ref{9}}) reduces to the
Gibbons equation (\textbf{\ref{6}}); the equation of the Riemann mapping (%
\textbf{\ref{8}}) reduces to the equation of the Riemann mapping (\textbf{%
\ref{5}}), where the generating function of the Miura type transformations is%
\begin{equation*}
p=C^{-1}q
\end{equation*}%
and the Miura type transformations are (see \textbf{\cite{Chen+Tu}})%
\begin{equation*}
B^{k}=C^{k}(C^{-1})^{k}\text{, \ \ \ \ }k=0,1,2,...
\end{equation*}

The Benney hydrodynamic chain is well known. Plenty hydrodynamic reductions
(see for instance, \textbf{\cite{Bogdan}}, \textbf{\cite{Krich}}, \textbf{%
\cite{Zakh}}) are found many years ago. Below we show links with
hydrodynamic reductions of hydrodynamic chains connected by the Miura type
and reciprocal transformations with the Benney hydrodynamic chain.

\section{Modified Benney hydrodynamic chain}

For instance, the Zakharov hydrodynamic reductions (see \textbf{\cite{Zakh}})%
\begin{equation}
a_{t}^{i}=\partial _{x}\left[ \frac{(a^{i})^{2}}{2}+A^{0}\right] \text{, \ \
\ \ \ }b_{t}^{i}=\partial _{x}(a^{i}b^{i})\text{, \ \ \ \ \ \ \ }A^{0}=%
\overset{N}{\underset{n=1}{\sum }}b^{n}  \label{neg}
\end{equation}%
of the Benney moment chain (\textbf{\ref{1}}) are connected with the
equation of the Riemann surface%
\begin{equation*}
\lambda =\mu +\overset{N}{\underset{n=1}{\sum }}\frac{b^{n}}{\mu -a^{n}}.
\end{equation*}%
Let us rewrite the above formula in the form%
\begin{equation}
\lambda =\mu +\frac{b^{i}}{\mu -a^{i}}+\underset{k\neq i}{\sum }\frac{b^{k}}{%
\mu -a^{k}},  \label{rim}
\end{equation}%
where the index $i$ is \textbf{fixed}. Substituting the Taylor series%
\begin{equation*}
\mu ^{(i)}=a^{i}+\frac{b^{i}}{\lambda }+\frac{c^{i}(\mathbf{a},\mathbf{b})}{%
\lambda ^{2}}+\frac{d^{i}(\mathbf{a},\mathbf{b})}{\lambda ^{3}}+...
\end{equation*}%
in (\textbf{\ref{zak}}), one can obtain an infinite series of conservation
laws at the vicinity of each puncture $\mu ^{(i)}=a^{i}$. The Benney
hydrodynamic chain (\textbf{\ref{cons}}) is determined for $k\geqslant 0$.
Let us extend this hydrodynamic chain in the \textit{negative} direction.
Then the negative part of the Benney hydrodynamic chain is given by%
\begin{equation*}
\partial _{t}\mathbf{H}_{-1}=\partial _{x}\left[ \frac{(\mathbf{H}_{-1})^{2}%
}{2}+\mathbf{H}_{0}\right] \text{, \ \ \ \ \ \ }\partial _{t}\mathbf{H}%
_{-k-1}=\frac{1}{2}\partial _{x}\left[ \overset{k}{\underset{m=0}{\sum }}%
\mathbf{H}_{-m-1}\mathbf{H}_{m-k-1}\right] \text{, \ \ \ \ }k=1,2,...,
\end{equation*}%
where $\mathbf{H}_{-1}\equiv a^{i}$, $\mathbf{H}_{-2}\equiv b^{i}$, and the
above hydrodynamic type system%
\begin{eqnarray*}
a_{t}^{k} &=&\partial _{x}\left( \frac{(a^{k})^{2}}{2}+\mathbf{H}_{-2}+%
\underset{n\neq i}{\sum }b^{n}\right) \text{, \ \ \ \ \ }b_{t}^{k}=\partial
_{x}(a^{k}b^{k})\text{, \ \ \ \ \ }k=1\text{, }2\text{, ... , }N\text{, \ \
\ \ \ }k\neq i, \\
&& \\
\partial _{t}\mathbf{H}_{-1} &=&\partial _{x}\left( \frac{(\mathbf{H}%
_{-1})^{2}}{2}+\mathbf{H}_{-2}+\underset{n\neq i}{\sum }b^{n}\right) \text{,
\ \ \ \ \ \ }\partial _{t}\mathbf{H}_{-2}=\partial _{x}(\mathbf{H}_{-1}%
\mathbf{H}_{-2})
\end{eqnarray*}%
is connected with the equation of the Riemann surface (see (\textbf{\ref{rim}%
}))%
\begin{equation}
\tilde{\lambda}\equiv \lambda ^{(i)}=\frac{\mathbf{H}_{-2}}{\mu -\mathbf{H}%
_{-1}}+\mu +\underset{k\neq i}{\sum }\frac{b^{k}}{\mu -a^{k}}.  \label{tot}
\end{equation}%
Since the first moment of the modified Benney chain $B^{0}\equiv \mathbf{H}%
_{-1}$, the equation of the Riemann surface for the corresponding
hydrodynamic reduction%
\begin{eqnarray*}
u_{t}^{k} &=&\partial _{x}\left( \frac{(u^{k})^{2}}{2}+B^{0}u^{k}\right) 
\text{,\ \ \ \ \ \ \ }b_{t}^{k}=\partial _{x}[(B^{0}+u^{k})b^{k}]\text{, \ \
\ \ \ }k=1\text{, }2\text{, ... , }N\text{, \ \ \ \ \ }k\neq i, \\
&& \\
B_{t}^{0} &=&\partial _{x}\left( \frac{(B^{0})^{2}}{2}+\mathbf{H}_{-2}+%
\underset{n\neq i}{\sum }b^{n}\right) \text{, \ \ \ \ }\partial _{t}\mathbf{H%
}_{-2}=\partial _{x}(B^{0}\mathbf{H}_{-2})
\end{eqnarray*}%
of the modified Benney hydrodynamic chain is given by%
\begin{equation*}
\lambda =\frac{\mathbf{H}_{-2}}{p}+B^{0}+p+\underset{k\neq i}{\sum }\frac{%
b^{k}}{p-u^{k}},
\end{equation*}%
where $u^{k}=a^{k}-B^{0}$; the generating function of the Miura type
transformations is $p=\mu -B^{0}$.

\section{Dispersionless 2+1 Harry Dym equation}

Finally, let us apply the reciprocal transformation (see (\textbf{\ref{soh}}%
))%
\begin{equation}
dz=\mathbf{H}_{-2}dx+B^{0}\mathbf{H}_{-2}dt\text{, \ \ \ \ \ \ }dy=dt
\label{ro}
\end{equation}%
to the above hydrodynamic type system.

Then the equation of the Riemann surface for the corresponding hydrodynamic
reduction%
\begin{eqnarray*}
\bar{u}_{y}^{k} &=&\partial _{z}\left[ \frac{(\bar{u}^{k})^{2}}{2}%
(C^{-1})^{2}\right] \text{,\ \ \ \ \ \ \ }\bar{b}_{y}^{k}=\partial
_{z}\left( C^{-1}\bar{u}^{k}\bar{b}^{k}\right) \text{, \ \ \ \ \ }k=1\text{, 
}2\text{, ... , }N\text{, \ \ \ \ \ }k\neq i, \\
&& \\
C_{y}^{0} &=&\left( 1+\underset{n\neq i}{\sum }\bar{b}^{n}\right)
C^{-1}C_{z}^{-1}+(C^{-1})^{2}\underset{n\neq i}{\sum }\bar{b}_{z}^{n}\text{,
\ \ \ \ \ }C_{y}^{-1}=(C^{-1})^{2}C_{z}^{0}
\end{eqnarray*}%
of a dispersionless limit of 2+1 Harry Dym equation is given by%
\begin{equation*}
\lambda =C^{-1}q+C^{0}+\frac{1}{q}+\underset{k\neq i}{\sum }\frac{\bar{b}^{k}%
}{q-\bar{u}^{k}},
\end{equation*}%
where $C^{0}\equiv B^{0}$, $C^{-1}\equiv \mathbf{H}_{-2}$, $\bar{b}%
^{k}=b^{k}/\mathbf{H}_{-2}$, $\bar{u}^{k}=u^{k}/\mathbf{H}_{-2}$; and the
generating function of the Miura type transformations is $p=qC^{-1}$.

Let us rewrite the equation of the Riemann surface (\textbf{\ref{rim}}) for
the Zakharov hydrodynamic reduction (\textbf{\ref{neg}}) in the form (see 
\textbf{\cite{Maks93}})%
\begin{equation*}
\lambda =p\underset{k=1}{\overset{N}{\prod }}\frac{p-c^{k}}{p-u^{k}},
\end{equation*}%
where $B^{0}\equiv \mathbf{H}_{-1}=\sum (u^{n}-c^{n})$. Application of the
reciprocal transformation (\textbf{\ref{ro}}) yields \textit{another}
hydrodynamic reduction of a dispersionless limit of 2+1 Harry Dym equation%
\begin{equation*}
\bar{u}_{y}^{k}=\partial _{x}\left[ \frac{(\bar{u}^{k})^{2}}{2}(C^{-1})^{2}%
\right] \text{, \ \ \ \ \ }\bar{c}_{y}^{k}=\partial _{x}\left[ \frac{(\bar{c}%
^{k})^{2}}{2}(C^{-1})^{2}\right] \text{, \ \ \ \ }k=1\text{, }2\text{, ..., }%
N,
\end{equation*}%
where\ $C^{-1}\equiv \mathbf{H}_{-2}=\prod (u^{n}/c^{n})\equiv \prod (\bar{u}%
^{n}/\bar{c}^{n})$. The corresponding equation of the Riemann surface is
given by%
\begin{equation*}
\lambda =q\underset{k=1}{\overset{N}{\prod }}\frac{1-q/\bar{c}^{k}}{1-q/%
\bar{u}^{k}}.
\end{equation*}

Let us start from the waterbag reduction (see \textbf{\cite{Gib+Yu}} and 
\textbf{\cite{Kodama}})%
\begin{equation*}
a_{t}^{k}=\partial _{x}\left[ \frac{(a^{k})^{2}}{2}+\underset{n=1}{\overset{N%
}{\sum }}\varepsilon _{n}a^{n}\right] \text{, \ \ \ \ \ }k=1\text{, }2\text{%
, ..., }N
\end{equation*}%
of the Benney moment chain (\textbf{\ref{1}})%
\begin{equation*}
\lambda =\mu -\varepsilon _{i}\ln (\mu -a^{i})-\underset{k\neq i}{\sum }%
\varepsilon _{k}\ln (\mu -a^{k})\text{, \ \ \ \ \ \ \ \ \ }\sum \varepsilon
_{n}=0.
\end{equation*}%
Let us identify $\mathbf{H}_{-1}\equiv a^{i}$, where the index $i$ is 
\textbf{fixed}. Then the above hydrodynamic type system%
\begin{eqnarray*}
a_{t}^{k} &=&\partial _{x}\left[ \frac{(a^{k})^{2}}{2}+\varepsilon _{i}%
\mathbf{H}_{-1}+\underset{n\neq i}{\sum }\varepsilon _{n}a^{n}\right] \text{%
, \ \ \ \ \ }k=1\text{, }2\text{, ... , }N\text{, \ \ \ \ \ }k\neq i, \\
&& \\
\partial _{t}\mathbf{H}_{-1} &=&\partial _{x}\left[ \frac{(A^{-1})^{2}}{2}%
+\varepsilon _{i}\mathbf{H}_{-1}+\underset{n\neq i}{\sum }\varepsilon
_{n}a^{n}\right]
\end{eqnarray*}%
is connected with the equation of the Riemann surface (cf. (\textbf{\ref{tot}%
}))%
\begin{equation*}
\tilde{\lambda}\equiv \lambda ^{(i)}=\frac{1}{\mu -\mathbf{H}_{-1}}\underset{%
k\neq i}{\prod }(\mu -a^{k})^{-\varepsilon _{k}/\varepsilon _{i}}\exp \frac{%
\mu }{\varepsilon _{i}}.
\end{equation*}%
Since the first moment of the modified Benney chain $B^{0}\equiv \mathbf{H}%
_{-1}$, the equation of the Riemann surface for the corresponding
hydrodynamic reduction%
\begin{eqnarray*}
u_{t}^{k} &=&\partial _{x}\left( \frac{(u^{k})^{2}}{2}+B^{0}u^{k}\right) 
\text{, \ \ \ \ \ }k=1\text{, }2\text{, ... , }N\text{, \ \ \ \ \ }k\neq i,
\\
&& \\
B_{t}^{0} &=&\partial _{x}\left( \frac{(B^{0})^{2}}{2}+\underset{n\neq i}{%
\sum }\varepsilon _{n}u^{n}\right) \text{,}
\end{eqnarray*}%
of the modified Benney chain is given by%
\begin{equation*}
\lambda =\frac{\mathbf{H}_{-2}}{p}e^{p/\varepsilon _{i}}\underset{k\neq i}{%
\prod }\left( 1-\frac{p}{u^{k}}\right) ^{-\varepsilon _{k}/\varepsilon _{i}}.
\end{equation*}%
Finally, let us apply the reciprocal transformation (\textbf{\ref{ro}}),
where%
\begin{equation*}
\mathbf{H}_{-2}=e^{B^{0}/\varepsilon _{i}}\underset{k\neq i}{\prod }%
(u^{k})^{-\varepsilon _{k}/\varepsilon _{i}}.
\end{equation*}%
Then the equation of the Riemann surface for the corresponding hydrodynamic
reduction%
\begin{eqnarray*}
\bar{u}_{y}^{k} &=&\partial _{z}\left[ \frac{(\bar{u}^{k})^{2}}{2}%
(C^{-1})^{2}\right] \text{, \ \ \ \ \ }k=1\text{, }2\text{, ... , }N\text{,
\ \ \ \ \ }k\neq i, \\
&& \\
C_{y}^{-1} &=&(C^{-1})^{2}\partial _{z}\left( \underset{n\neq i}{\sum }%
\varepsilon _{n}\ln \bar{u}^{n}\right) \text{,}
\end{eqnarray*}%
of the dispersionless limit of 2+1 Harry Dym equation is given by%
\begin{equation*}
\lambda =C^{-1}q-\underset{n\neq i}{\sum }\varepsilon _{n}\ln \frac{1-\bar{u}%
^{n}/q}{\bar{u}^{n}}.
\end{equation*}%
All other hydrodynamic reductions of the Benney moment chain can be
re-calculated into hydrodynamic reductions of (\textbf{\ref{g}}) in the same
way.

\section*{Acknowledgement}

I am grateful to the Institute of Mathematics in Taipei (Taiwan) where some
part of this work has been done, and especially to Jen-Hsu Chang and Derchyi
Wu for fruitful and stimulating discussions.%

\end{document}